\documentclass[twocolumn,showpacs,preprintnumbers,amsmath,amssymb]{revtex4}
\usepackage{graphicx}% Include figure files
\usepackage{dcolumn}% Align table columns on decimal point
\usepackage{bm}% bold math
\usepackage{epsfig}
\usepackage{amsfonts}

\begin{document}

%\preprint{APS/123-QED}

\title{From time series to superstatistics}% Force line breaks with \\

\author{Christian Beck}
\affiliation{School of Mathematical Sciences, Queen Mary,
University of London, Mile End Road, London E1 4NS, UK}

\author{Ezechiel G.D. Cohen}
\affiliation{The Rockefeller University, 1230 York Avenue, New
York, New York 10021, USA}

\author{Harry L. Swinney}
\affiliation{Center for Nonlinear Dynamics and Department of
Physics, University of Texas at Austin, Austin, Texas 78712, USA}

\date{\today}

\begin{abstract}
Complex nonequilibrium systems are often effectively described by a
`statistics of a statistics', in short, a `superstatistics'. We
describe how to proceed from a given experimental time series to a
superstatistical description. We argue that many experimental data
fall into three different universality classes:
$\chi^2$-superstatistics (Tsallis statistics), inverse
$\chi^2$-superstatistics, and log-normal superstatistics. We discuss
how to extract the two relevant well separated superstatistical time
scales $\tau$ and $T$, the probability density of the superstatistical
parameter $\beta$, and the correlation function for $\beta$ from the
experimental data.  We illustrate our approach by applying it to
velocity time series measured in turbulent Taylor-Couette flow, which
is well described by log-normal superstatistics and exhibits clear
time scale separation.

\end{abstract}

%Use showkeys class option if keyword
\maketitle

\section{Introduction}

Driven nonequilibrium systems of sufficient complexity are often
effectively described by a superposition of different dynamics on
different time scales. As a simple example consider a Brownian
particle moving through a changing environment. A relatively fast
dynamics is given by the velocity of the Brownian particle, and a
slow dynamics is given, e.g., by temperature changes of the
environment. The two effects are associated with two well
separated time scales, which result in a superposition of two
statistics, or in a short, a superstatistics (SS)
\cite{beck-cohen, boltzmann-m, touchette-beck, grigolini,
plastino, prl, souza, chavanis, luczka, beck03, reynolds, boden,
beck-physica-d, jung-swinney, daniels, rapisarda, rap2, cosmic,
maya, abe-turner, hasegawa, bouchard, ausloos}. The stationary
distributions of superstatistical systems typically exhibit
non-Gaussian behavior with fat tails, which can decay with a power
law, or as a stretched exponential, or in an even more complicated
way. An essential ingredient of SS models is the existence of an
intensive parameter $\beta$ that fluctuates on a large
spatio-temporal scale $T$. For the above example of a Brownian
particle, $\beta$ is the fluctuating inverse temperature of the
environment, but in general $\beta$ can also be an effective
friction constant, a changing mass parameter, a changing amplitude
of Gaussian white noise, the fluctuating energy dissipation in
turbulent flows, or simply a local variance parameter extracted
from a signal.
%If $\beta$ is distributed
%according to a particular probability distribution, the
%$\chi^2$-distribution, then the corresponding marginal
%distributions, obtained by integrating over all $\beta$, are
%given by the generalized canonical distributions of nonextensive
%statistical mechanics
%\cite{tsa1,tsa2,tsa3,abe,souza,souza2,chavanis}. For other
%distributions of the intensive parameter $\beta$, one ends up
%with more general statistics, which contain Tsallis statistics as
%a special case.
The SS concept is quite general and has recently been applied to a
variety of physical systems, including Lagrangian~\cite{reynolds,
beck03, boden} and Eulerian~\cite{beck-physica-d, jung-swinney}
turbulence, defect turbulence~\cite{daniels}, atmospheric
turbulence~\cite{rapisarda, rap2}, cosmic ray
statistics~\cite{cosmic}, solar flares~\cite{maya}, random
networks~\cite{abe-turner, hasegawa} and mathematical
finance~\cite{bouchard, ausloos}.

In this paper we address a problem that is of great interest in
experimental applications. Given an experimentally measured time
series or signal, how can we check if this time series is well
described by a superstatistical model, i.e., does  it contain two
separate time scales, and how can we extract the relevant
superstatistical parameters from the time series? Further, since
there are infinitely many SS \cite{beck-cohen}, which ones are the
most relevant for typical experimental situations?
%We will provide a general recipee to tackle this
%problem. First, we will develop the theoretical foundations.

We argue that many experimental data (see, e.g., \cite{rapisarda,
reynolds, jung-swinney, sattin, boden, castaing, BLS}) are well
described by three major universality classes, namely $\chi^2$,
inverse $\chi^2$, and log-normal SS. These SS represent a
universal limit statistics for large classes of dynamical systems.
We then show how to extract the superstatistical
parameters from a given experimental signal. Our example is a time
series of longitudinal velocity differences measured in turbulent
Taylor-Couette flow \cite{lewis-swinney}. We will extract the two
relevant time scales $\tau$ and $T$ from the data and show that
there is clear time scale separation for our example. We will also
investigate the probability density of $\beta$ and the
$\beta$-correlation function. While our turbulent time series
appears to fall into the universality class of log-normal
superstatistics, our concepts are general and can in
principle be applied to any time series.

\section{Superstatistical universality classes}
Consider a driven nonequilibrium system which is inhomogeneous and
consists of many spatial cells with different values of some
intensive parameter $\beta$ (e.g., the inverse temperature). The
cell size can be determined by the correlation length of the
continuously varying $\beta$-field. Each cell is assumed to reach
local equilibrium very fast, i.e., the associated relaxation time
$\tau$ is short. 
%A test particle stays for the rather long time
%scale $T$ in each cell, and during this time $\beta$ that is
%approximately constant.
The parameter $\beta$ in each cell is approximately
constant during the time scale $T$, then it changes to a new value.
In the long-term run $(t>>T)$, the
stationary distributions of this inhomogeneous system arise as a
superposition of Boltzmann factors $e^{-\beta E}$ weighted with
the probability density $f(\beta)$ to observe some value $\beta$
in a randomly chosen cell:
\begin{equation}
p(E)=\int_0^\infty f(\beta)  \frac{1}{Z(\beta)} \rho(E) e^{-\beta
E}d\beta  \label{ppp}
\end{equation}
Here $E$ is an effective energy for
each cell, $\rho(E)$ is the density of
states, and $Z(\beta)$ is the normalization constant of $\rho(E)
e^{-\beta E}$ for a given $\beta$.
The simplest example is a Brownian particle of mass $m$
moving through a changing environment in $d$ dimensions. For its
velocity $\vec{v}$ one has the local Langevin equation
\begin{equation}
\dot{\vec{v}}=-\gamma \vec{v} + \sigma \vec{L}(t) \label{lange}
\end{equation}
($\vec{L}(t)$: $d$-dimensional Gaussian white noise) which becomes
superstatistical because for a fluctuating
environment the parameter $\beta :=\frac{2}{m}
\frac{\gamma}{\sigma^2}$ becomes a random variable as well: it
varies from cell to cell on the large spatio-temporal scale $T$.
Of course, for this example $E=\frac{1}{2}m\vec{v}^2$, and while
on the time scale $T$ the local stationary distribution in each
cell is Gaussian,
\begin{equation}
p(\vec{v}|\beta)=\left( \frac{\beta}{2\pi}\right)^{d/2}
e^{-\frac{1}{2}\beta m\vec{v}^2},
\end{equation}
the marginal distribution describing the long-time
behavior of the particle for $t>>T$,
\begin{equation}
p(\vec{v})=\int_0^\infty f(\beta)p(\vec{v}|\beta)d\beta
\label{margi}
\end{equation}
exhibits non-trivial behavior. The large-$|v|$ tails of the
distribution (\ref{margi}) depend on the behavior of $f(\beta )$
for $\beta \to 0$ \cite{touchette-beck}. For example, if
$f(\beta)$ is a $\chi^2$-distribution of degree $n$, then
eq.~(\ref{margi}) generates Tsallis statistics \cite{tsa1, abe},
with entropic index $q$ given by $q=1+\frac{2}{n+d}$ \cite{prl}.
Of course, a necessary condition for a superstatistical
description to make sense is the condition $\tau = \gamma^{-1}
<<T$, because otherwise the system is not able to reach local
equilibrium before the next change of $\beta$ takes place. In
superstatistical turbulence models \cite{prl, reynolds, beck03,
beck-physica-d, boden}, one formally replaces the variable
$\vec{v}$ in eq.~(\ref{lange}) by the velocity difference
$\vec{u}$ (or acceleration $\vec{a}$ on smallest scales), and
$\beta$ is related to energy dissipation $\epsilon$.

The distribution $f(\beta)$ is determined by the spatio-temporal
dynamics of the entire system under consideration. By
construction, $\beta$ is positive, so $f(\beta)$ cannot be
Gaussian. Let us here consider three examples of what to expect in
typical experimental situations.

(a) There may be many (nearly) independent microscopic random
variables $\xi_j$, $j=1,\ldots , J$, contributing to
$\beta$ in an additive way. For large $J$ their rescaled sum
$\frac{1}{\sqrt{J}}\sum_{j=1}^J\xi_j$ will approach a Gaussian
random variable $X_1$ due to the Central Limit Theorem. In total,
there can be many different random variables consisting of
microscopic random variables, i.e., we have $n$ Gaussian random
variables $X_1,\ldots ,X_n$ due to various relevant degrees of
freedom in the system. As mentioned before, $\beta$ needs to be
positive; a positive $\beta$ is obtained by squaring these
Gaussian random variables. The resulting $\beta=\sum_{i=1}^nX_i^2$
is $\chi^2$-distributed with degree $n$,
\begin{equation}
f(\beta )=\frac 1{\Gamma (\frac n2)}\left( \frac n{2\beta _0}\right)
^{n/2}\beta ^{n/2-1}e^{-\frac{n\beta }{2\beta _0}},  \label{chi2}
\end{equation}
where $\beta_0$ is the average of $\beta$. As shown in \cite{wilk,
prl}, the SS resulting from (\ref{margi}) and (\ref{chi2}) is
Tsallis statistics \cite{tsa1}. It exhibits power-law tails for
large $|\vec{v}|$. Our above argument shows that Tsallis
statistics arises as a universal limit dynamics, i.e., the details
of the microscopic random variables $\xi_j$ (e.g., their
probability densities) are irrelevant.

(b) The same considerations as above can be applied if the
'temperature' $\beta^{-1}$ rather than $\beta$ itself is the sum
of several squared Gaussian random variables arising out of many
microscopic degrees of freedom $\xi_j$. The resulting $f(\beta)$
is the inverse $\chi^2$-distribution given by
\begin{equation}
f(\beta )=\frac{\beta _0}{\Gamma (\frac n2)}\left( \frac{n\beta _0}2\right)
^{n/2}\beta ^{-n/2-2}e^{-\frac{n\beta _0}{2\beta }}. \label{chi2inv}
\end{equation}
It generates superstatistical
distributions (\ref{margi})
that have exponential decays in $|\vec{v}|$
\cite{sattin, touchette-beck}.
Again this superstatistics is universal: details of the $\xi_j$
are irrelevant.

(c) Instead of $\beta$ being a sum of many contributions, for
other systems (in particular, turbulent ones) the random variable
$\beta$ may be generated by multiplicative random processes. We
may have a local cascade random variable $X_1= \prod_{j=1}^{J}
\xi_j$, where $J$ is the number of cascade steps and the $\xi_j$
are positive microscopic random variables. By the Central Limit
Theorem, for large $J$ the random variable $\frac{1}{\sqrt{J}}
\log X_1= \frac{1}{\sqrt{J}} \sum_{j=1}^J \log \xi_j$ becomes
Gaussian for large $J$. Hence $X_1$ is log-normally distributed.
In general there may be $n$ such product contributions to $\beta$,
i.e., $\beta = \prod_{i=1}^n X_i$. Then $\log \beta = \sum_{i=1}^n
\log X_i$ is a sum of Gaussian random variables; hence it is
Gaussian as well. Thus $\beta$ is log-normally distributed, i.e.,
\begin{equation}
f(\beta )=\frac{1}{\sqrt{2\pi}s\beta}
\exp \left\{ \frac{-(\ln \frac{\beta}{\mu})^2}{2s^2}\right\},
\label{logno}
\end{equation}
where $\mu$ and $s^2$ are suitable mean and variance parameters
\cite{beck-cohen}.
For related turbulence models, see, e.g., \cite{kolmo, castaing,
beck03, reynolds}. Again this log-normal result is independent of
the details of the microscopic cascade random variables $\xi_j$;
hence log-normal SS is universal as well.

Although more complicated cases can be constructed, we believe
that most experimentally relevant cases fall into one of these
three universality classes, or simple combinations of them.
$\chi^2$ superstatistics has been found for wind velocity
fluctuations \cite{rapisarda, rap2}, and log-normal
superstatistics has been found for Lagrangian \cite{beck03,
reynolds, boden} and Eulerian \cite{beck-physica-d, jung-swinney}
turbulence. Candidate systems for inverse $\chi^2$ superstatistics
are systems exhibiting velocity distributions with exponential
tails \cite{touchette-beck, sattin}.

%In the following we restrict ourselves to the simplest types of
%superstatistical models, those produced by a superposition of
%Gaussian distributions with a varying local variance $\beta^{-1}$
%and average 0.

\section{Application to experimental time series}

Suppose some
experimental time series $u(t)$ is given
\cite{kantz}. Our goal is to test the
hypothesis that it is due to a superstatistics and if so, to
extract the two basic time scales
$\tau$ and $T$ as well as $f(\beta)$. First let us
determine the large time scale $T$. For this we divide
the time series into $N$ equal time intervals of size $\Delta t$. The
total length of the signal is $t_{max}=N\Delta t$. We then define
a function $\kappa(\Delta t)$ by
\begin{equation}
\kappa (\Delta t) = \int_0^{t_{max}-\Delta t} dt_0
\frac{\langle (u-\bar{u})^4 \rangle_{t_0,\Delta t}}{\langle
(u-\bar{u})^2 \rangle^2_{t_0,\Delta t}} \label{kappa}
\end{equation}
Here $\langle \cdots \rangle_{t_0,\Delta t}=\int_{t_0}^{t_0+\Delta t}\cdots
dt$ denotes an integration over an interval of length
$\Delta t$ starting at $t_0$, and $\bar{u}$
is the average of $u(t)$ (we may either
choose $\bar{u}$ to be a local average in each cell or a global average
over the entire time series ---
our results do not depend on this choice in a significant way).
Equation~(\ref{kappa}) simply means that
the local flatness is evaluated in each interval of length $\Delta t$,
and the result is then averaged over all $t_0$.
We now
define the superstatistical time scale $T$ by the condition
\begin{equation}
\kappa (T)= 3. \label{3}
\end{equation}
Clearly this condition simply implies that we are looking for the
simplest SS, a superposition of local Gaussians, which have local
flatness 3 (see \cite{jung-swinney} for similar ideas). If $\Delta
t$ is so small that only one constant value of $u$ is observed in
each interval, then of course $\kappa(\Delta t)=1$. On the other
hand, if $\Delta t$ is so large that it includes the entire time
series, then we obtain the flatness of the distribution of the
entire signal, which will be larger than 3, since superstatistical
distributions are fat-tailed. Hence there exists a time scale $T$
satisfying (\ref{3}).

The function $\kappa (\Delta t)$ is shown in Fig.~1 for
longitudinal velocity differences, $u(t)=v(t+\delta) -v(t)$,
measured in Taylor-Couette flow at Reynolds number $Re
=540000$~\cite{lewis-swinney}.  The total number of measurement
points was $2\times 10^7$, and in the present analysis $\Delta t \leq 1000$,
so $N\geq 2 \times 10^4$, which means there is sufficient statistics 
to obtain
precise values for the time scales $T$ and $\tau$. 
\begin{figure}
\epsfig{file=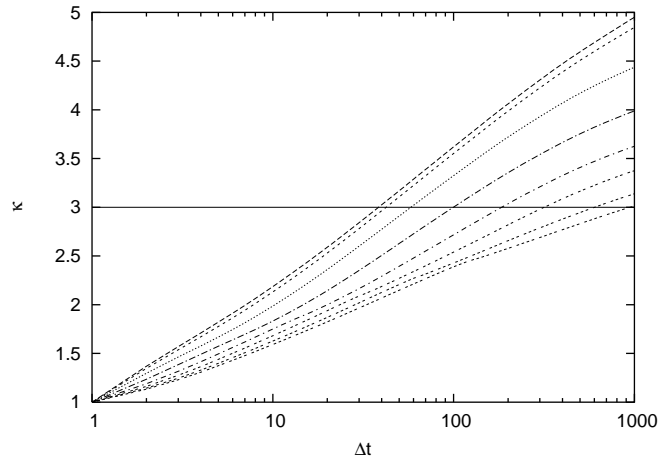} \caption{Determination of the
superstatistics long time scale $T$ from the flatness function
$\kappa (\Delta t )$ given in eq. (8), for $\delta =2^j,
j=0,1,2,\ldots ,7$ (from top to bottom). The intersections with
the line $\kappa =3$ yield $T=39, 42, 58, 100, 184, 320, 600,
948$, respectively. Time is nondimensional (see text).}
\end{figure}
For details of the experiment, see \cite{lewis-swinney}. For each
time difference $\delta$ 
(measured in units of the sampling frequency, which was 2500 times
the inner cylinder rotation frequency), the relevant
superstatistical time scale $T$ leading to locally Gaussian
behavior is extracted in Fig.~1. The time scales $T$ have to be
compared with the relaxation times $\tau =\gamma^{-1}$ of the
dynamics, which can be estimated from the short-time exponential
decay of the correlation function $C_u(t-t')= \langle
u(t)u(t')\rangle$ of the velocity difference $u$ (Fig. 2).
\begin{figure}
\epsfig{file=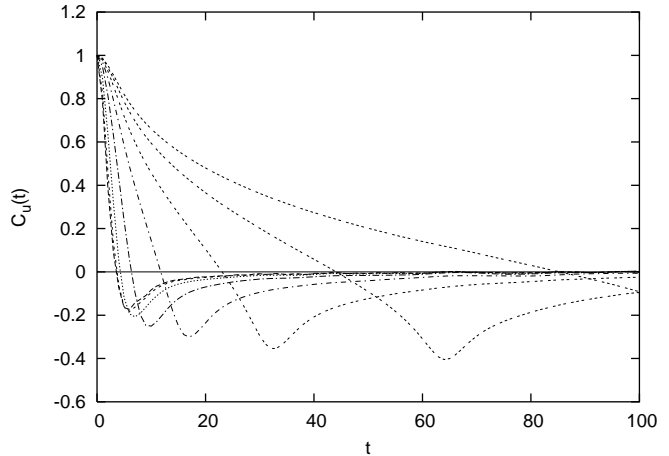} \caption{Determination of the
superstatistics short time scale $\tau$ from the decay of the
correlation function $C_u(t)$ of the velocity difference.
Defining $\tau$ by $C_u(\tau)=e^{-1}C_u(0)$, we obtain for
$\delta=2^j,j=0,1,2,\cdots,7$,
$\tau=2.1,2.3,2.8,4.3,7.2,12.1,19.9,29.5$, respectively.}
\end{figure}
We find that the ratio $T/\tau \approx 17...34$ is large compared
to unity, and the ratio has only a weak (logarithmic) dependence
on $\delta$ (Fig.~3). Thus there are indeed two well separated
time scales in the time series for turbulent Couette-Taylor flow.
\begin{figure}
\epsfig{file=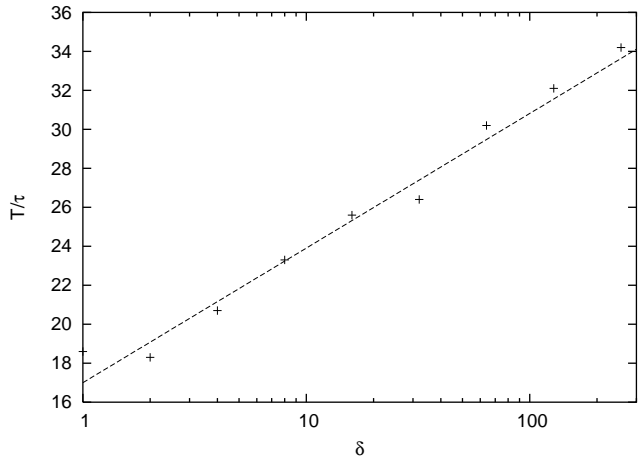} \caption{The time scale ratio
$T/\tau$, given as a function of $\delta$ for turbulence data at
$Re=540000$, is large compared to unity. Thus the long and short
time scales are well-separated, as required for superstatistics.
The dashed line is a fit given by $T/\tau =17+3\ln \delta$.}
\end{figure}

Laboratory data were obtained for a wide range of Reynolds
numbers, so we can also examine how the time scale ratio changes
with the Reynolds number $Re$. Fig.~4 shows that $T/\tau$
increases with increasing $Re$, meaning that the superstatistics
approach becomes more and more exact for $Re \to \infty$.
\begin{figure}
\epsfig{file=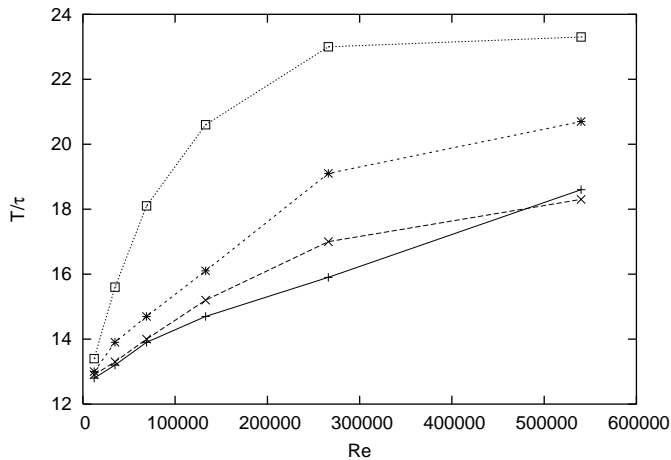} \caption{Time scale ratio
$T/\tau$ as a function of Reynolds number $Re$ for $\delta
=8,4,2,1$ (from top to bottom).}
\end{figure}

Next, we are interested in the analysis of the slowly varying stochastic
process $\beta (t)$. Since the variance of the local Gaussians
$\sqrt{\frac{\beta}{2\pi}} e^{-\frac{1}{2}\beta u^2}$
is given by $\beta^{-1}$, we can determine the process $\beta (t)$
from the time series as
\begin{equation}
\beta (t_0) = \frac{1}{\langle u^2 \rangle_{t_0,T}-\langle u \rangle^2_{t_0,T}}
.
\end{equation}
We obtain the probability density $f(\beta)$ as a histogram of
$\beta (t_0)$ for all values of $t_0$, as shown in Fig.~5. 
We compare the experimental data with
log-normal, $\chi^2$ and inverse $\chi^2$ distributions with the
same mean $\langle \beta \rangle$ and variance $\langle
\beta^2\rangle -\langle \beta \rangle^2$ as the experimental data.
The fit of the data to a log-normal distribution is significantly
better than to a $\chi^2$ or inverse $\chi^2$. Indeed, the cascade
picture of energy dissipation in turbulent flows suggests that our
time series should belong to the log-normal universality class of
superstatistics (see section II (c)). A power-law relation between
the energy dissipation rate $\epsilon$ and $\beta$ was found for
the Couette-Taylor turbulence data in an analysis in
\cite{jung-swinney}. Note that if such a power-law relation is
valid, then
a log-normally distributed $\epsilon$ implies a log-normally distributed
$\beta$, and vice versa.

\begin{figure}
\hspace*{7cm} {\bf (a)} \epsfig{file=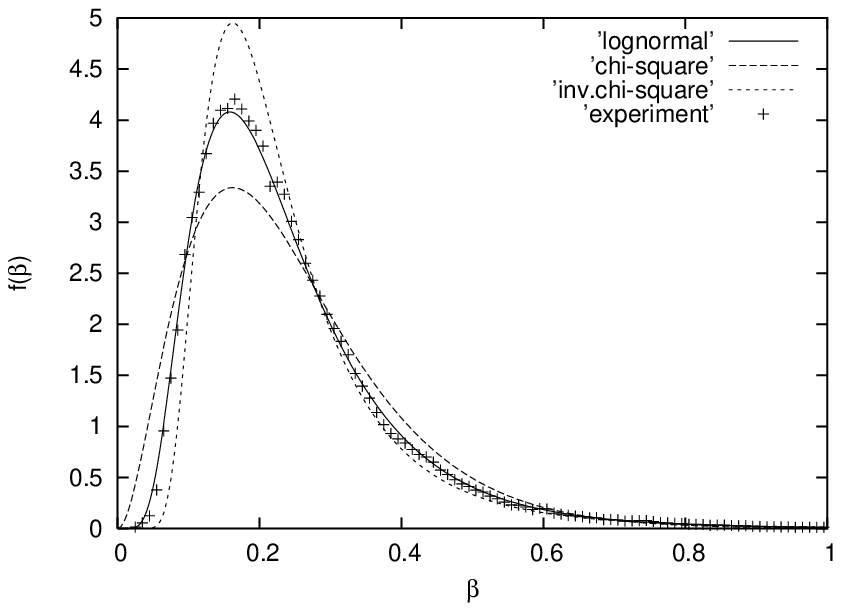}
\hspace*{7cm} {\bf (b)} \epsfig{file=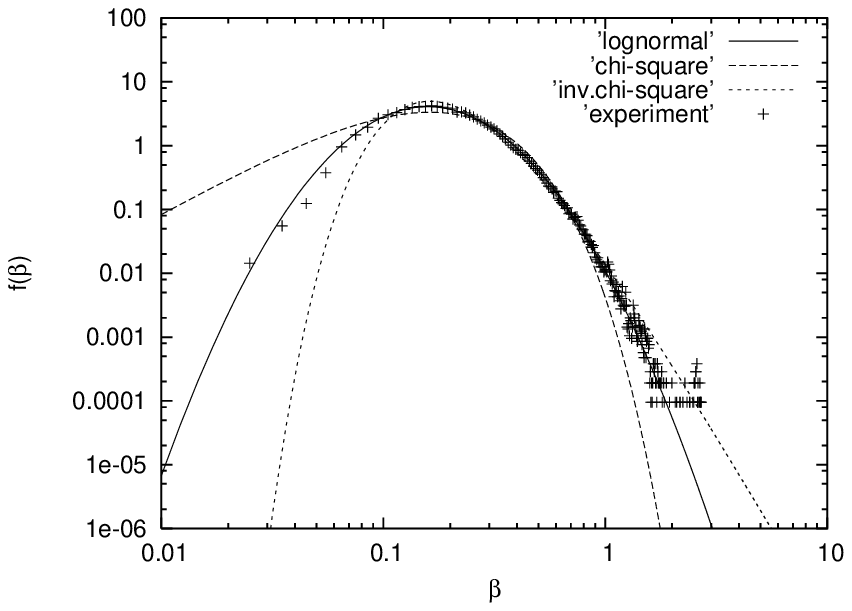}
\caption{Probability density
$f(\beta)$ extracted from the turbulent time series $(\delta=16)$,
and compared with log-normal, $\chi^2$, and inverse $\chi^2$
distributions, on (a) linear-linear and (b) log-log plots. All
distributions have the same mean and variance as the experimental
data.}
\end{figure}

%\begin{figure}
%$\epsfig{file=supertime-Fig5.eps} \caption{Probability density
%$f(\beta)$ as extracted from the turbulent time series
%$(\delta=16)$, and comparison with a log-normal, $\chi^2$, and
%inverse $\chi^2$ distribution, all having the same mean and
%variance as the experimental data.}
%\end{figure}
%\begin{figure}
%\epsfig{file=supertime-Fig6.eps}
%\caption{Same as Fig.~5, but a double logarithmic plot
%is chosen. This emphasizes the tails.}
%\end{figure}

For superstatistics to make physical sense, the variable $\beta$
must change slowly compared to $u$. This is indeed the case for
our turbulence data, as Fig.~6 illustrates for a sample time
series.
\begin{figure}
\epsfig{file=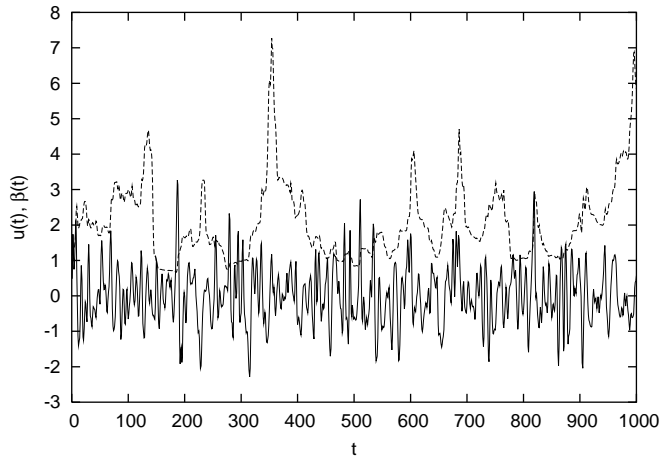}
\caption{Time series of $\beta(t)$ (top)
and $u(t)$ (bottom). For this example $\delta =2$
and $T=42$.}
\end{figure}
A slow $\beta$-dynamics also implies a slow correlation decay of
the $\beta$-correlation function $C_\beta (t-t')= \langle \beta
(t) \beta (t') \rangle$. For our data we observe a power-law decay
with a small exponent, $C_\beta(t)\sim t ^{-0.9}$ (Fig.~7).
\begin{figure}
\epsfig{file=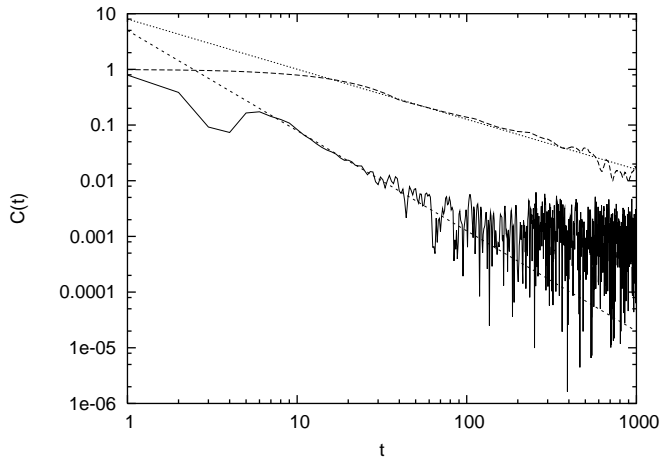} \caption{Correlation functions
$C_\beta (t)$ (top) and $|C_u(t)|$ (bottom) for $\delta =1$. The
straight lines represent power laws with exponents -0.9 and -1.8.}
\end{figure}
%\footnote{Similar power-law decays are observed for the
%correlation function of volatility in
%financial time series \cite{bouchard}.}
This means that $\beta(t)$ indeed has a long memory and changes
slowly, a necessary consistency condition for the superstatistics
approach. For comparison, the correlation function of the
longitudinal velocity difference $u(t)$ first decays exponentially
fast and only finally, for very large times, approaches a
power-law decay of the form $C_u(t) \sim t^{-1.8}$, as shown in
Fig.~7. Note that for $t>6$ the $\beta$-correlation is larger than
the $u$-correlation by a factor 10...100.

Finally, we may check the validity of the general
SS formula
\begin{equation}
p(u)=\int_0^\infty f(\beta) p(u|\beta) d\beta , \label{pre}
\end{equation}
where $p(u|\beta)$ is the conditional distribution of the signal
$u(t)$ in cells of size $T$, and $p(u)$ is the marginal stationary
distribution of the entire signal. For log-normal superstatistics
this means
\begin{equation}
p(u)=\frac{1}{2\pi s} \int_0^\infty d\beta \beta^{-1/2}
\exp \left\{ \frac{-(\ln (\beta /\mu))^2}{2s^2}\right\}
e^{-\frac{1}{2}\beta u^2}. \label{lognoss}
\end{equation}
As shown in Fig.~8, there is excellent
agreement between the experimental histogram and the
superstatistical model prediction, both in the tails and in the
region around the peak of the distribution.

\begin{figure}
\hspace*{7cm} {\bf (a)} \epsfig{file=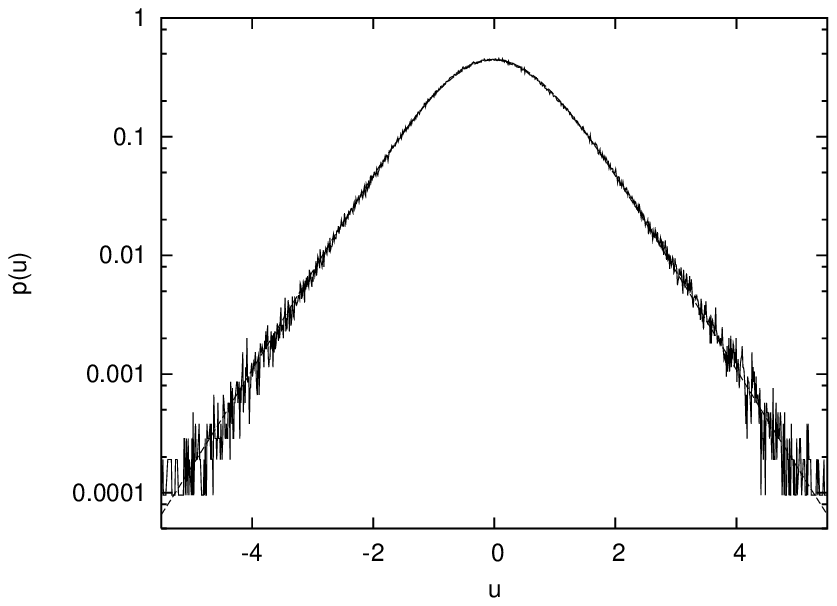}
\hspace*{7cm} {\bf (b)} \epsfig{file=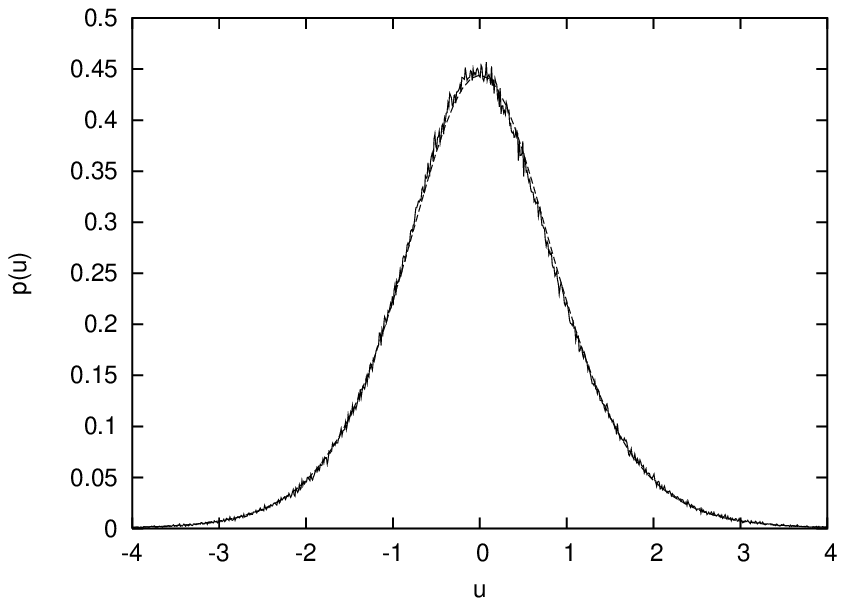} 
\caption{Comparison of the measured (fluctuating lines) and
predicted (dashed lines) probability distribution $p(u)$ for
velocity differences on (a) semi-log plot, which emphasizes the
tails, and (b) linear-linear plot, which emphasizes the peak. The
predicted $p(u)$ was obtained from eq.~(\ref{lognoss}) with
$f(\beta)$ being a log-normal distribution with the same
parameters as in Fig.~5 $(\delta =16)$.}
\end{figure}

%\begin{figure}
%\epsfig{file=supertime-Fig10.eps}
%\caption{Same as Fig.~9 but a linear scale is chosen. This emphasizes
%the vicinity of the maximum.}
%\end{figure}

For any superstatistics one can formally define a parameter $q$ by
\cite{beck-cohen}
\begin{equation}
q:= \frac{\langle \beta^2 \rangle}{
\langle \beta \rangle^2} . \label{13}
\end{equation}
$q$ measures in a quantitative way the deviation from Gaussian
statistics. No fluctuations in $\beta$ at all correspond to
$f(\beta)=\delta (\beta -\beta_0)$ and $q=1$, i.e., ordinary
equilibrium statistical mechanics.
% (i.e.\
%no fluctuations in $\beta$ at all,
%$f(\beta)=\delta
%(\beta -\beta_0)$).
For $\chi^2$-superstatistics, the above $q$ is given by $q=1+2/n$
\cite{beck-cohen, wilk} and is nearly the same as the entropic
index $q_{T}$ introduced by Tsallis \cite{tsa1, abe, prl}
($q_{T}=1+2/(n+1)$). For log-normal (LN) superstatistics, one can
relate $q$ to the flatness $F=\frac{\langle u^4 \rangle}{\langle
u^2 \rangle^2}$ of the distribution $p(u)$. 
Since for log-normal
superstatistics~\cite{beck-physica-d}
\begin{eqnarray}
\langle \beta  \rangle_{LN} &=& \mu e^{\frac{1}{2} s^2}\\
\langle \beta^2 \rangle_{LN} &=& \mu^2 e^{2s^2}\\
\langle u^2 \rangle &=&\mu^{-1}e^{ \frac{1}{2} s^2} \\
\langle u^4 \rangle &=& 3 \mu^{-2} e^{2s^2}, \label{lasteq}
\end{eqnarray}
where $\mu$ and $s^2$ are the mean and variance parameters
of the distribution (\ref{logno}), we
arrive at the following simple relation,
\begin{equation}
q=e^{s^2}=\frac{1}{3} F,  \label{14}
\end{equation}
using eq.~(\ref{13}).
We thus have two different equations to evaluate $q$ for our time
series. The first one, based on eq.~(\ref{13}), is always valid
(i.e., for any $f(\beta)$), whereas the second one, based on
eq.~(\ref{14}), should coincide with the first one provided the
system is described by log-normal superstatistics. Figure~9 shows
the $q$-values that we extract from the experimental data for
$Re=540000$ using both methods.
\begin{figure}
\epsfig{file=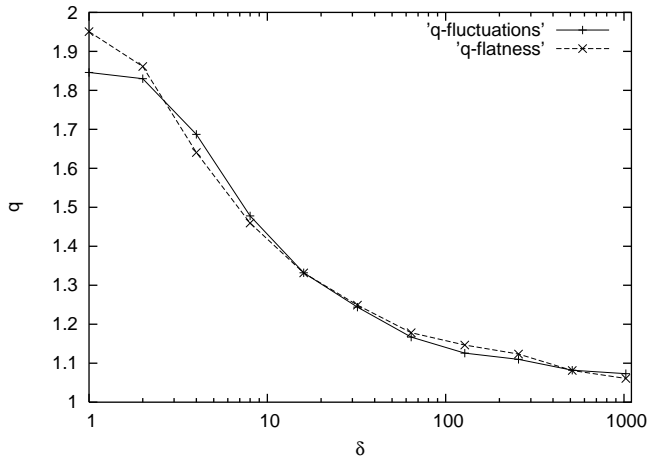} \caption{The parameter $q$ (eq.
(13)) as a function of $\delta$, as evaluated from eq.~(\ref{13})
(q-fluctuations) and eq.~(\ref{14}) (q-flatness).}
\end{figure}
As expected, $q$ decreases monotonically with scale $\delta$. Both
curves agree well for $\delta \geq 2$. This indicates that
log-normal superstatistics is a good model for our data, and that
our extraction of the time scale $T$ of the process $\beta (t)$ is
consistent. Significant deviations between the two $q$-values
occur only on the smallest scale $\delta =1$, where the experiment
reaches its resolution limits.

\section{Conclusions}

In this paper we have advocated the view that the non-Gaussian
behavior of many complex driven nonequilibrium systems can often be
understood as a superposition of two different statistics on different
time scales, in short, a superstatistics. We have argued that typical
experimental situations are described by three relevant universality
classes, namely $\chi^2$, inverse $\chi^2$, and log-normal
superstatistics. Our example, turbulent Taylor Couette flow, falls
into the universality class of log-normal superstatistics. This means
the time series is essentially described by local Boltzmann factors
$e^{-\frac{1}{2}\beta u^2}$ whose variance parameter $\beta$ varies
slowly according to a log-normal distribution function. We have
developed a general method to extract from data the process $\beta
(t)$, its probability density $f(\beta)$, and its correlation
function. Our approach is applicable to any experimental time
series. We have extracted the two relevant time scales $\tau$ (the
relaxation time to local equilibrium) and $T$ (the large time scale on
which the intensive parameter $\beta$ fluctuates).
%Our main example was an experimentally
%measured time series of longitudinal velocity differences
%in a turbulent Couette-Taylor flow.
%Based on the very accurate data of Lewis and Swinney \cite{lewis-swinney},
%we show that
%the data can be very well described by a simple superstatistical model
%which falls into the log-normal universality class.
Our main result is that for turbulent Taylor-Couette flow there is
clear time scale separation, which is a necessary condition for a
superstatistical description to make physical sense. The ratio
$T/\tau$
%extracted from the data is indeed large and it
grows logarithmically with the scale separation $\delta$
on which longitudinal
velocity differences are investigated. Moreover, $T/\tau$ also
increases with increasing Reynolds number, making the
superstatistical approach more and more exact for increasing
Reynolds number. The experimentally measured distributions of
$\beta$ and $u$ agree
very well with the superstatistical model predictions.

\end{document}